\documentclass[%
 aip,
cp,  
 amsmath,amssymb,
 reprint,%
nofootinbib,%
]{revtex4-2}

\usepackage[dvipsnames]{xcolor}
\usepackage{graphicx}
\usepackage{dcolumn}
\usepackage{bm}

\usepackage[utf8]{inputenc}
\usepackage[T1]{fontenc}
\usepackage{mathptmx} 
\usepackage{siunitx} 
\usepackage{gensymb} 
\usepackage[normalem]{ulem} 

\usepackage[caption=false]{subfig}

\graphicspath{{figs/}}

\newcommand{\geant}{{\textsc{Geant4}}}

\newcommand{\mwe}{\text{m\,w.\,e.}}
\newcommand{\mwee}{\text{m\,w.\,e}}
\newcommand{\mkwe}{\text{km\,w.\,e.}}

\begin{document}

\title{Muon-induced background in a next-generation dark matter experiment based on liquid xenon}

\author{Viktor P\v{e}\v{c}} 
 \email[Corresponding author: ]{viktor.pec@fzu.cz}
\affiliation{
 FZU -- Institute of Physics of the Czech Academy of Sciences, Na Slovance 1999/2, 182 21 Prague, Czech Republic $^{\rm{i)}}$ \\
}
\author{Vitaly A. Kudryavtsev}%
 \email{v.kudryavtsev@sheffield.ac.uk}
\affiliation{%
Department of Physics and Astronomy, University of Sheffield, Sheffield S3 7RH, United Kingdom
}%

\date{\today} 

\begin{abstract}
      Muon-induced neutrons can lead to potentially irreducible
      backgrounds in rare event search experiments. We have investigated
      the implication of laboratory depth on the muon induced background
      in a future dark matter experiment capable of reaching the so-called
      neutrino floor. Our simulation study focuses on a xenon-based
      detector with 71~tonnes of active mass, surrounded by additional
      veto systems including an instrumented water shield. Two locations at the Boulby
      Underground Laboratory (UK) served as a case study: an experimental
      cavern in salt at a depth of 2850~\mwe{} (similar to the location of
      the existing laboratory), and a deeper laboratory located in
      polyhalite rock at a depth of 3575~\mwee. Our results show that less
      than one event of cosmogenic background is likely to survive
      standard analysis cuts for 10 years of operation at either
      location. The largest background component that we identified comes
      from delayed neutron emission from $^{17}$N which is produced from
      $^{19}$F in the fluoropolymer components of the experiment. Our
      results confirm that a dark matter search with sensitivity to the
      neutrino floor is viable (from the point of view of cosmogenic
      backgrounds) in underground laboratories at these levels of rock
      overburden. We present details of the performed simulations
      and of the obtained results.
\end{abstract}

\maketitle
\renewcommand{\thefootnote}{\roman{footnote})}
\footnotetext{Previously at the University of Sheffield}

\section{\label{sec:intro}Introduction}

The choice of underground laboratory for a future high-sensitivity experiment for rare-event searches will be heavily dependent on the depth of the site, as events triggered by cosmic-ray muons can constitute significant background to the signal events. For example, in searches for dark matter (DM) Weakly Interacting Massive Particles (WIMP), isolated neutrons originating in the muon activity can mimic dark matter interaction if scattering only once in the active region producing a single nuclear recoil (NR). We focused our attention on a potential next-generation DM experiment based on Liquid Xenon Time Projection Chamber (LXe-TPC) technology, and we investigated whether the depth of about 3\,\mkwe{} was sufficient for such experiment.

The work was performed as a part of feasibility study for Boulby Mine (UK) to host the next-generation LXe-TPC experiment \cite{HA2021} and we wanted to learn whether the current underground laboratory in Boulby was deep enough regarding the cosmic-ray induced backgrounds. A new larger site would have to be excavated nearby the existing laboratory which is at 1100\,m level underground, or 2850\,\mwe, and it is located within a layer of salt (NaCl). We also considered a new site at a deeper location at a 1400\,m level, or 3575\,\mwe{} underground. The Boulby mine served as a case model, however, the results are relevant to sites of similar depth.

We have also compared our \geant{} results on neutron production in simplified-geometry simulations for various materials and muon energies with other simulations and experimental data. Some results and distributions of interest are included in Ref.~\cite{pec2022}.

\section{\label{sec:sim}Simulation of cosmic-ray muons}

\subsection{Geometry}

We carried out simulations to determine the rate of potential background events caused by cosmic-ray muons in a next generation dark matter experiment with design based on the design of LUX-ZEPLIN (LZ) detector \cite{LZexp2019}. The main detector is a dual-phase xenon time projection chamber (hereafter LXe-TPC) containing 70~tonnes of active liquid xenon (LXe), corresponding to a $\sim$10-fold upscale of existing experiments LZ \cite{Aalbers2022} and XENONnT \cite{Aprile2022}.

\begin{figure}[!htb]
    \centering
    \subfloat[]{
        \includegraphics[width=0.45\textwidth]{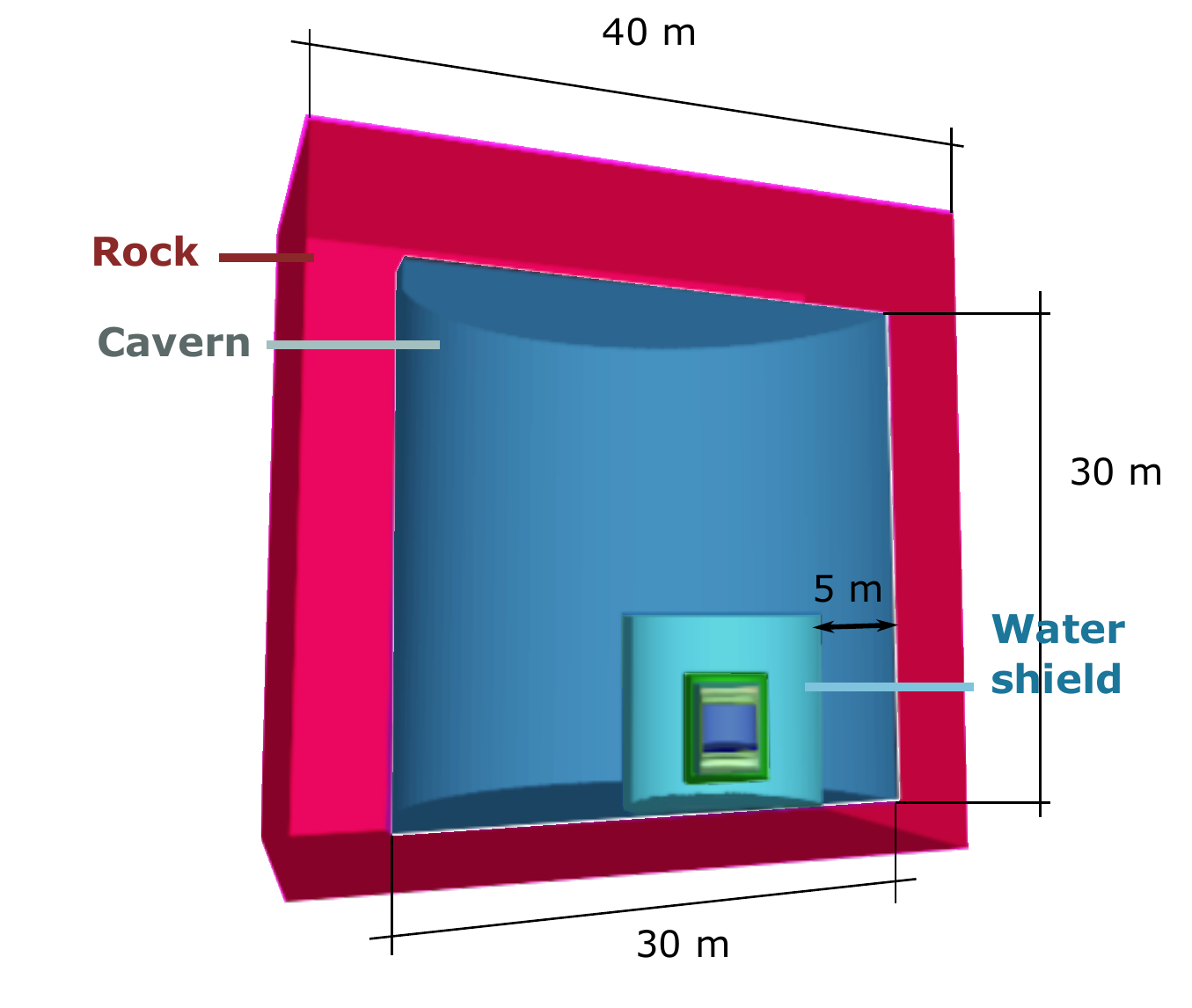}
        \label{fig:geom:a}
    }
    \subfloat[]{
        \includegraphics[width=0.55\textwidth]{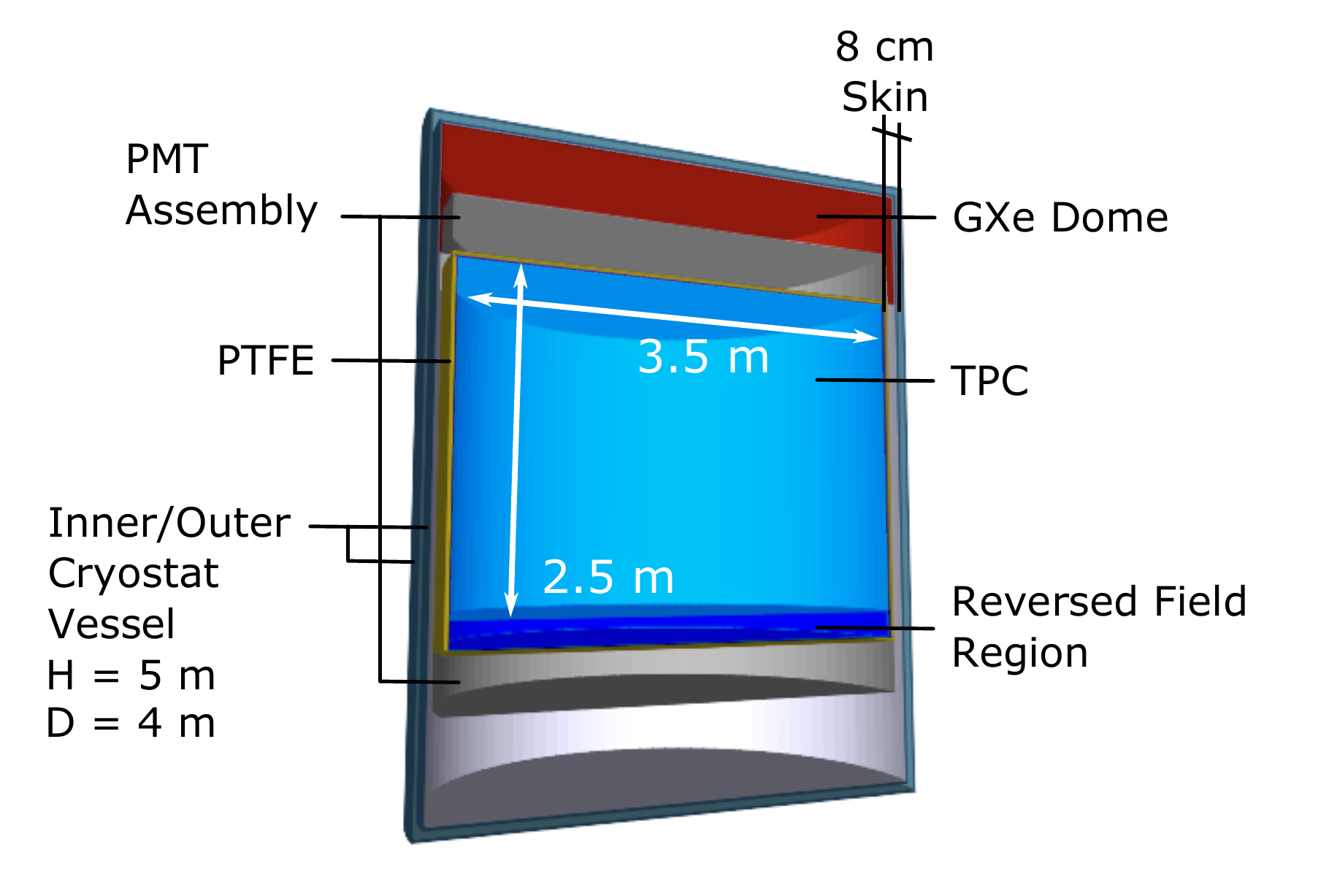}
        \label{fig:geom:b}
    }

    \caption{Illustrations of the geometry model used in the simulations. \protect\subref{fig:geom:a} Cross-sectional view of the cavern (bluish grey) within the rock (red) with the water shield (light blue) holding the scintillator (green) and the detector cryostat. \protect\subref{fig:geom:b} Cross-sectional view of the cryostat with the main LXe-TPC (blue).}
    \label{fig:geom}
\end{figure}

We implemented a simplified detector geometry model with the main elements included: a vacuum cryostat approximately 4\,m in diameter and 5\,m in height containing the xenon detector, with an anti-coincidence veto system made of gadolinium-loaded liquid scintillator (GdLS), 50\,cm thick, surrounding the cryostat, and all submerged in a water tank with 12\,m in diameter and 11\,m in height for shielding from local radioactivity backgrounds. The water tank was placed into a cylindrical cavern of diameter and height of 30\,m. The cavern was surrounded by rock material with at least 5\,m on the sides and 7\,m and 3\,m at the top and the bottom, respectively. These dimensions of the rock material were motivated by secondary production along the muon propagation in the rock, as discussed in the next section. We tested two different rock compositions, one with salt (NaCl, \SI{2.17}{g\per \cm\cubed}) corresponding to the current laboratory location in Boulby mine, and one with polyhalite (K$_2$Ca$_2$Mg(SO$_4$)$_4${}$\cdot$2H$_2$O, \SI{2.78}{g\per \cm\cubed}) corresponding to the potential new site. Overview of the geometry can be seen in the cross-sectional view of the cavern in the illustration in Fig.~\ref{fig:geom:a}.

The central part of the cryostat was the active LXe-TPC, cylindrical, 3.5~m and 2.5~m in diameter and height. There was an 8\,cm thick layer of LXe around the active volume separated by a PTFE field cage, which will be called LXe skin hereafter. Arrays of photo-multiplier tubes (PMT) meant to read the fast scintillation light from LXe and delayed electroluminescence signal from Xe gas (GXe) of the TPC were approximated with uniform cylinders made of steel with reduced density of \SI{0.4}{g/cm^3}, or about 5$\%$ of the standard density of steel, simulating metal components of the structure of the arrays and matching its mass. Illustration of a cross-sectional view of the cryostat is shown in Fig.~\ref{fig:geom:b}.

The water tank, GdLS and LXe skin volumes were assumed to be instrumented with photo-sensitive detectors and each of them would serve as a veto system by registering either Cherenkov light in the case of the water tank, or scintillation light in the case of GdLS and LXe skin.

\subsection{Underground muons}

Cosmic-ray muons were sampled on the top and side surfaces of a 40\,m cube that surrounded the cavern such that they needed to travel through at least 7\,m of rock at the top of the cavern and through at least 5\,m of rock on the sides. Production of high-energy cascades and fast neutrons in the rock that could propagate into the cavern was expected to reach equilibrium with their absorption within that distance. 

Distributions of primary energies and directions were calculated using the MUSIC and MUSUN codes \cite{music1997,music} (Ref.~\cite{music} describes the procedure and muon transport through rock down to experimental site). The mean muon energy and zenith angle were calculated as \SI{261}{GeV} (\SI{134}{GeV} median) and 30.6\degree{} (30.1\degree{} median), respectively. The rate of simulated muons was \SI{0.8759}{s^{-1}} for the existing Boulby site within salt at 2850\,\mwe\ vertical overburden. This rate was determined based on the measured flux in the existing laboratory, \SI{3.75\pm0.09e-8}{cm^{-2}s^{-1}}~\cite{lr2013}. The surface profile was assumed to be flat in these simulations (in reality, variations in elevation up to 30\,m exist on the surface over areas of a few km$^2$). For the proposed deeper site in polyhalite at 3575\,\mwe\ vertical overburden, the same muon distributions were used, but the equivalent sampling rate was recalculated to be \SI{0.2625}{s^{-1}}.

We used \geant{} version 10.5 simulation toolkit to simulate the transport of muons through the modelled experimental site. Physical processes were modelled according to the toolkit's modular physics list Shielding. In total, 800~million muons were simulated for each rock material, salt and polyhalite. 
These numbers correspond to approximately 29\,years and 97\,years of live time of the experiment, respectively.

\section{\label{sec:ana}Neutron background analysis}

The expected WIMP signature in a typical xenon-based dark matter experiment consists of a single scatter event at low energy, usually $\lesssim$50\,keV, which is classified as a nuclear recoil (NR) using specific discrimination techniques, and which is in anti-coincidence with veto systems. We adopted a simple background counting technique with the potential (irreducible) background identified as a single NR of energy above 1\,keV.

We analysed energy depositions in active regions of the experimental setup which were stored in the simulation output. The detector response to the depositions was not simulated (i.e.~the digitised PMT waveforms resulting from the fast scintillation light from LXe and delayed electroluminescence signal from GXe from each energy deposition in the active volume) and was considered only in terms of the characteristic times over which signals were collected and the equivalent energy thresholds in the respective active volumes -- LXe-TPC, LXe skin, liquid scintillator, water tank. We selected events with depositions above 1\,keV from a single NR at least 5\,cm from the boundary of the active region in 1\,ms readout time window of the LXe-TPC. We required there was no other NR recoil above 0.5\,keV, which we deemed resolvable in the delayed signal from GXe. We also required there were no non-NR depositions above 10\,keV in total, including a quenching factor of 4 for protons and heavier ionising particles.

Energy depositions in the skin, liquid scintillator and water tank were summed over \SI{1}{\micro\second}, irrespective of their origin. We chose this time window to emulate the shaping time of the photosensitive instrumentation of the systems. We chose thresholds of 100\,keV, 200\,keV and 200\,MeV in the skin, liquid scintillator and water tank, respectively, to trigger veto signals. These thresholds were chosen based on our previous experience with detector modelling and the existing water Cherenkov experiments, mainly their energy thresholds and photo-coverage. We consider them to be conservative. The selected events in the LXe-TPC were then tested for anti-coincidence with the veto signals by requiring no veto signal to be present within 0.5\,ms before or after any TPC signal.

\begin{figure}[!htb]
    \subfloat[]{
        \includegraphics[width=0.49\textwidth]{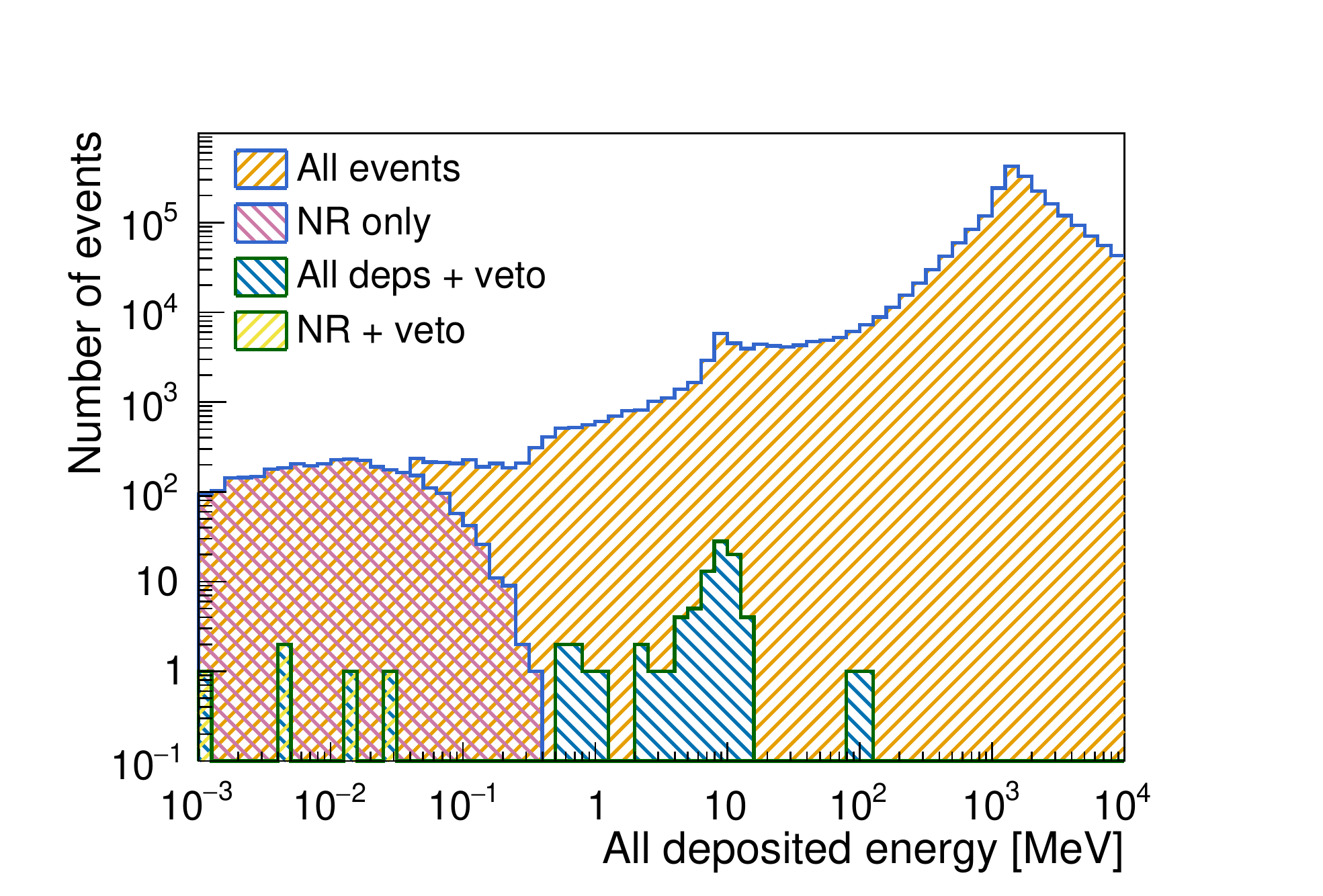}
        \label{fig:deps:a}
    }
    \subfloat[]{
        \includegraphics[width=0.49\textwidth]{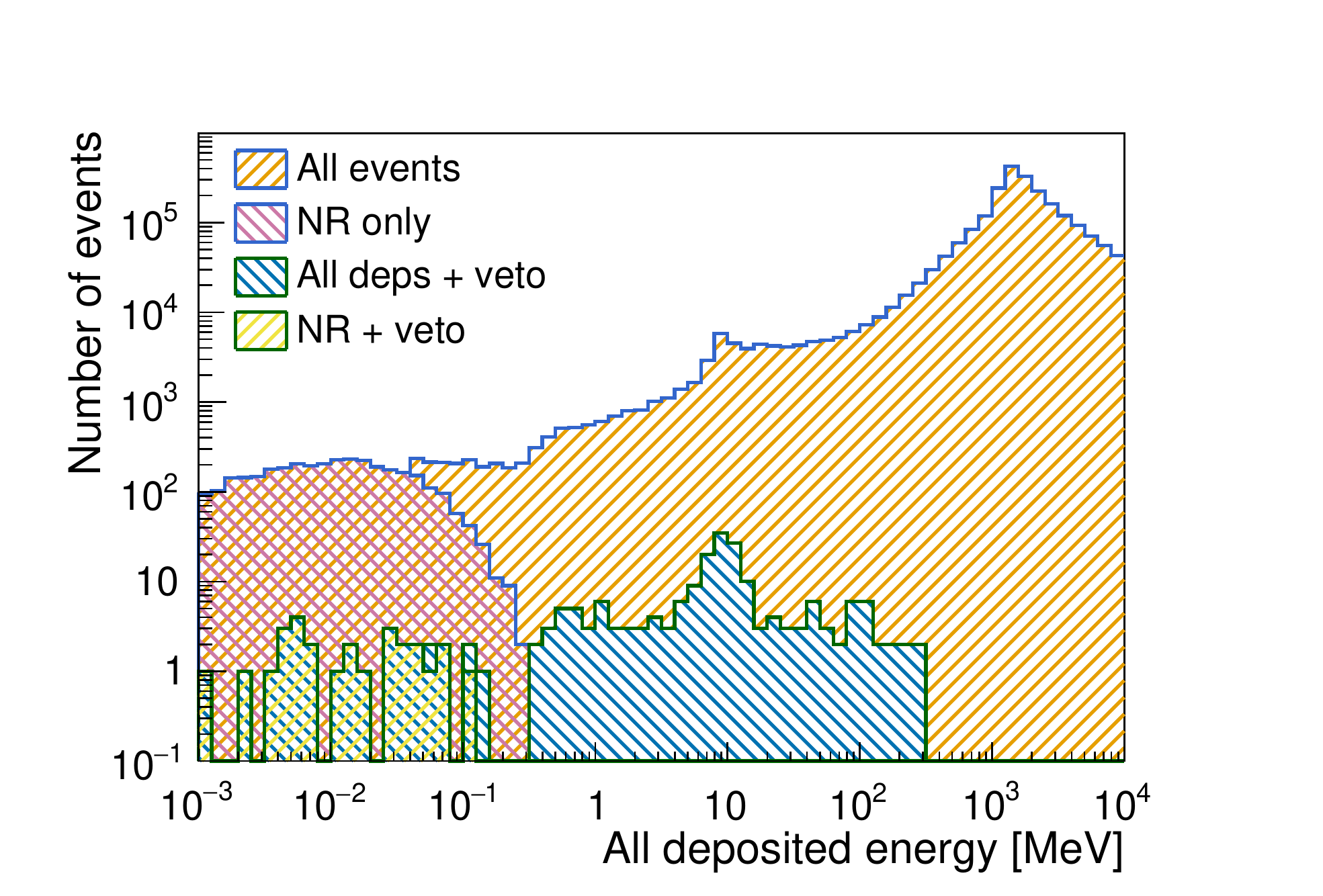}
        \label{fig:deps:b}
    }
    \caption{Energy depositions in the active region of the LXe-TPC for NaCl as the rock material. Events with any NR activity were considered. \protect\subref{fig:deps:a} Full configuration containing the GdLS. \protect\subref{fig:deps:b} Configuration where GdLS was considered to be part of the water shield.}
    \label{fig:deps}
\end{figure}

Spectra of depositions in the LXe-TPC at various stages of the selection are shown in Fig.~\ref{fig:deps:a}. Depositions in all events with some NR are compared to events where only NR happened. One can see the low-energy part ($<$100\,keV) of all the events is dominated by depositions from NR. The same distributions are also shown after application of the veto condition. Recognized activity near the active region of the TPC (in the LXe skin, GdLS, or the water tank) together with the requirement on only NR to be observed reduces the number of events by a factor \num{e6}.

We also wanted to answer the question whether the scintillator veto was necessary for reduction of the cosmogenic background. We reused the simulated samples and emulated the geometry without the GdLS by assuming all depositions within its volume were part of the water tank, and applied the corresponding 200\,MeV threshold on the depositions from the combined volume. Fig.~\ref{fig:deps:b} shows comparison of deposition spectra the same way is in Fig.~\ref{fig:deps:a} but this time the GdLS was not included in the veto. The number of events in the LXe-TPC passing the veto is larger than in the case with the GdLS. We must note here that these are events before the requirement on multiplicity of NR was applied.

After application of the full selection process described above, including the rejection of multiple scatters and exclusion of 5\,cm of LXe from all sides of the TPC (64\,t fiducial volume), no events passed the selection in the sample with NaCl as the material of the surrounding rock. A single event passed the selection in the sample with polyhalite. As a result, the estimated background event rate in the fiducial volume of the LXe-TPC (64\,t) is $<$0.84\,events (90\%\,CL) in 10 years for the 2850\,\mwe~site in salt.  For the 3578\,\mwe~site in polyhalite the rate is estimated to be 0.10 (0.01--0.45 at 90\%\,CL) events in 10 years of running.

In the case where no liquid scintillator is used as an additional veto system, no events were observed for the site in salt, and the total of 2 events were observed for the site in polyhalite. The respective estimated background rates correspond to $<$0.84\,events~(90\%\,CL) in 10\,years (2850\,\mwe~site, salt), and 0.31~(0.11--0.77 at 90\%\,CL) events in 10\,years (3575\,\mwe~site, polyhalite). The limits and confidence intervals are statistical only. The estimated rates in both cases are well below the expected background of tens of events in 10\,years extrapolated from the estimates in Table~III of Ref.~\cite{Akerib2020}. 

The observed single nuclear recoils are located at the boundary of the fiducial volume. Most events with isolated neutrons causing NRs in the LXe-TPC and avoiding the veto were coming from delayed neutron emission from the PTFE field cage. Secondary particles in muon-induced cascade activated fluorine producing $^{17}$N from $^{19}$F. $^{17}$N undergoes $\beta$ decay with half-life of 4.2\,s with emission of a neutron.

We also made considerations of systematic uncertainties. We concluded that the major uncertainty was due to the neutron production modelling in \geant{}. From comparisons with previous studies (see, for instance, Refs.~\cite{al2009,lr2013} for discussions of simulations and comparison of different models with experimental data) we conservatively assessed the uncertainty to be a factor of 2. Uncertainties in the muon flux are linked to an unknown location of the new cavern within the existing level. Our estimate of 10\% was based on measurements at different locations. We estimated the uncertainty of 20\% for the deeper site where the flux was calculated based on the geophysical model of the Boulby mine but the exact location has not been determined. We approximated the muon spectrum at the deeper site to be the same as the one at the current laboratory. However, the calculated mean muon energies differ by about 9\% (261\,GeV and 282\,GeV for the 2850\,\mwe\ and 3575\,\mwe\ sites, respectively) and we expect it would lead to a small increase in the neutron production yield of (6--7)\%. This change is small compared with the other mentioned systematic uncertainties.

\section{\label{sec:concl}Conclusion}

We have performed simulations for an experiment similar in design to the LZ detector, upscaled to 71\,tonnes of LXe acting as a target. We conclude that, after applying a standard simplified analysis procedure and cuts, the event rate caused by cosmogenic activity stays below 1 event per 10\,years in the fiducial volume of the LXe-TPC (64\,tonnes). This rate is well below the expected background of tens of events from beta decays from radon progeny and ERs/NRs from physics backgrounds such as two-neutrino double beta decay of $^{136}$Xe and solar/atmospheric neutrinos with ER events leaking into NR band due to limited discrimination. The low rate also allows the experiment to reach the so-called neutrino floor, background from coherent scattering off of a nucleus of neutrinos from various sources (the Sun, cosmogenic and diffuse supernova neutrinos). From the point of view of cosmogenic background, the depth of about 3\,k\mwe\ or deeper is sufficient for a next generation dark matter experiment based on liquid xenon. Although there are significant systematic uncertainties related mainly to the modeling of neutron production, they cannot change our conclusion. The observed residual background of NR events comes from the production and delayed $\beta-n$ decay of $^{17}$N in PTFE (on fluorine) where only a single neutron scatter is detected. Our material budget contained about \SI{2.8}{\tonne} of PTFE. Although the residual background is very low, the design of a future experiment may need to limit PTFE usage to the necessary minimum. 

We have also investigated two designs of veto system: a default one with instrumented liquid scintillator surrounding the cryostat, and an option without the scintillator. No significant difference was observed between the two scenarios which lead us to the conclusion that the additional veto system is not required to suppress cosmogenic backgrounds for the sensitivity at the studied depth. This conclusion, however, does not apply to other types of backgrounds (mainly from detector components) where liquid scintillator is particularly efficient in tagging neutron-induced events. The decision about the need for a scintillator veto can only be taken after a detailed simulation of all types of background from all components is completed.

\begin{acknowledgments}
We wish to thank UKRI-STFC for financial support of the Boulby case study and the whole team who worked on the Boulby Feasibility Study project.
\end{acknowledgments}

\nocite{*}
\bibliography{bibfile}

\end{document}